# Precision game engineering through reshaping strategic payoffs

Elie Eshoa[1,2,3,4] and Ali R. Zomorrodi[3,4]*

[1] Computer Science Department, Harvard John A. Paulson School of Engineering and Applied Sciences, Boston, MA, USA
[2] Harvard Kenneth C. Griffin Graduate School of Arts and Sciences, Cambridge, MA, USA
[3] Mucosal Immunology and Biology Research Center, Pediatrics Department, Massachusetts General Hospital, Boston, MA, USA
[4] Harvard Medical School, Boston, MA, USA

* Corresponding author
Ali R. Zomorrodi
azomorrodi@mgh.harvard.edu

# Abstract

Nash equilibrium is a key concept in game theory fundamental for elucidating the equilibrium state of strategic interactions, finding applications in diverse fields such as economics, political science, and biology. However, the Nash equilibrium may not always align with the optimal or desired outcomes within a system. This article introduces a novel game engineering framework that tweaks strategic payoffs within a game to achieve a desired Nash equilibrium while averting undesired ones. Leveraging mixed-integer linear programming, this framework identifies intricate combinations of players and strategies and optimal perturbations to their payoffs that enable the shift from undesirable Nash equilibria to more favorable ones. We demonstrate the effectiveness and scalability of our approach on games of varying complexity, ranging from simple prototype games such as the Prisoner's Dilemma and Snowdrift games with two or more players to complex game configurations with as high as $10^6$ entries in the payoff matrix. These studies showcase the capability of this framework in efficiently identifying the alternative ways of reshaping strategic payoffs to secure desired Nash equilibria and preclude the undesired equilibrium states. Our game engineering framework offers a versatile toolkit for precision strategic decision-making with far-reaching implications across diverse domains.

# Introduction

In a mathematical game involving strategic interactions between two or more players, the Nash equilibrium (NE) is a state where no player can improve their payoff by changing their chosen strategy, provided the other players keep their strategies unchanged. In other words, the strategy of each player is optimal given the strategies of other players. This concept provides a powerful analytical tool to infer the outcome of strategic interactions amongst several decision-makers and aids in analyzing the long-term patterns of behaviors that emerge in multiple-player games. NE has been widely applied and adapted in diverse fields ranging from political and behavioral sciences [1, 2] to sociology [3], economics [4, 5], and biology [6-9]. An example of its use in political science is the concept of Mutually Assured Destruction (MAD), which describes a national security and military policy where neither of two nuclear-armed nations has any incentive to initiate aggression as they would otherwise suffer from complete destruction due to the use of nuclear arms [10]. In economics, NE has been used, for example, to predict the behavior of firms in an oligopoly (a market structure in which a small number of firms dominate and control the market) and the resulting market outcomes [11, 12]. In the field of biology, NE has been used to study interactions at all scales including among different biomolecules [6], different cell types (such as cooperation and conflict among microbes in a microbial community) [7, 8], different viruses [13, 14], as well as interactions between host and microbes or viruses [9, 15, 16], or between animals [17]. Overall, NE has been established as a powerful means of delineating the outcome of complex strategic interactions across a wide range of applications.

A situation that arises frequently in many real-world applications is that the NE of a game may not align with the desired or optimal outcome within a system. This is particularly important as the NE does not necessarily represent an optimal strategy with respect to the entire system. For example, in a Prisoner's Dilemma microbial game, the NE, which involves two (or more) microbial players defecting (e.g., by not synthesizing a metabolite essential for their growth that is costly to synthesize), will lead to the collapse of the entire microbial population. Likewise, MAD is an NE, which is highly undesired as it would lead to the total annihilation of all parties involved; this played a key role in preventing a full-scale nuclear war between the nuclear-armed states during the Cold War. In such cases, we have a strong tendency to move away from the undesired NE and establish another equilibrium, which is more favorable.

Prior studies investigating the attainment of a new NE have explored the concept of controllability, which is the ability of a system to be driven from an initial state to a desired state by the influence of a regulator. For instance, Zhang and Guo [18] derived sufficient and necessary conditions for a game-based control system. The controllability of NE has been examined in various systems, such as non-cooperative dynamical and adaptive games, which can exhibit both time-varying and time-independent characteristics [18-20]. To assess the controllability of NE, these studies utilized variational methods. These methods involve analyzing the dynamic equations representing the system's temporal behavior, while accounting for the payoffs of multiple agents, to determine the necessary and sufficient conditions under which the system can transition from an initial NE to a desired NE [18, 21].

Although considerable progress has been made in determining the feasibility of attaining a new NE, there is a dearth of research on methodologies to drive the system toward an equilibrium of interest. In this article, we aim to address this critical gap by taking a distinct approach to changing the NE of a game. More specifically, instead of focusing on the controllability of a game, here we propose to implement targeted interventions within the system. These interventions are precisely designed to alter the strategic payoffs of interacting agents, thereby incentivizing them to adopt strategies that converge to a more desirable outcome (NE). We refer to this approach as "game engineering" in this article. We tackled this by developing a novel optimization-based game engineering framework. The efficacy and scalability of our approach were validated by applying it to games of varying complexity. Through these proof-of-concept case studies, we demonstrate the capacity of our game engineering framework to effectively and robustly tweak strategic payoffs and interactions towards more desirable equilibrium states.

## Results

We aimed to design targeted interventions within a game that precisely modify the strategic payoffs of players to steer from an existing NE to a new and more desired one. To achieve this, we developed an automated game engineering framework based on Mixed-Integer Linear Programming (MILP), which adeptly identifies intricate combinations of payoff adjustments that would guarantee a seamless transition of the NE to a desired state, while precluding undesired NE. In essence, our pipeline meticulously pinpoints specific players, strategies, and their respective payoffs to tweak and the optional extent of perturbations to payoffs, all while minimizing overall changes to the payoff matrix (see Methods and **Figure 1**). Moreover, our tool can enumerate diverse and unique ways of reshaping the payoffs that would lead to the NE we seek to achieve while preventing the undesired NE. It is worth noting that, throughout this article, by NE, we mean pure strategy Nash equilibrium, where each player chooses one specific strategy at equilibrium (not a probability distribution over several strategies like in mixed strategies NE). In the following, we demonstrate the utility of our game engineering approach on prototype games with progressively increasing complexity. A sample implementation of the games discussed in the following sections is available as a Google Collaboratory notebook in **Supplementary File 1**.

**A Prisoner's Dilemma game with a 2-by-2 payoff matrix**: Consider a simple two-player game where each player can choose from one of two strategies, namely, to Cooperate (C) or Defect (D) with the payoffs shown in **Figure 2a**. This setting represents a classic Prisoner's Dilemma, where cooperation is rewarded, but betrayal can be tempting, thus both players end up defecting, i.e., (D, D) is the NE—an outcome we seek to avoid. Our goal here is to engineer the game so as to have (C, C) as the new NE while precluding (D, D) from being an NE. Our game engineering pipeline identifies eight alternative ways of achieving this goal as depicted in **Figures 2b** to **2i**. It is worth noting that in finding these engineering strategies we prevented our optimization formulation from perturbing the payoffs of the cell corresponding to the original NE, (D, D). This is with the assumption that it is practically more feasible to incentivize alternative strategies in real-world situations in order to render the current undesired NE unattainable. This constraint can be readily removed if desired.

The first identified game engineering strategy includes the following changes to the payoff matrix (**Figure 2b**): (i) Decreasing the payoff of defection for Player 1 when Player 2 cooperates, i.e., in cell (D, C) of the payoff matrix, from 0 to -1.01. This will incentivize Player 1 to cooperate, instead of defecting when Player 2 cooperates (payoff of -1 for cooperation vs. -1.01 for defection). (ii) Decreasing the payoff of defection for Player 2 when facing a cooperating Player 1, i.e., in cell (D, C) of the payoff matrix, from 0 to -1.01. This motivates Player 2 to cooperate when Player 1 cooperates (payoff of -1 for cooperation vs. -1.01 for defection). These two perturbations establish (C, C) as the new NE. (iii) Lastly, increasing the payoff of cooperating for Player 2 when facing a defecting Player 1, i.e., in cell (D, C) of the payoff matrix, from -4 to -2.99. This makes it more desirable for Player 2 to cooperate instead of defecting when Player 1 defects (payoff of -2.99 for cooperation vs. -3 for defection), thereby precluding (D, D) from being an NE.

We also uncovered seven additional ways of tweaking the payoffs to successfully transition from (D, D) to (C, C) as the NE (**Figures 2c** to **2i**). For example, an alternative path to achieve this, which also yields the same minimum total perturbations to the payoff matrix, involves perturbing three payoffs in two cells by positive values (**Figure 2c**). These include increasing the payoffs of Players 1 and 2 when they both cooperate, (C, C), and the payoff of cooperation for Player 2 when Player 1 defects, i.e., in cell (D, C). The former makes cooperation a favorable strategy for each player when the other one cooperates, while the latter makes cooperation more desirable for Player 2 when Player 1 defects, preventing (D, D) from being an NE.

Taken together, we were able to identify eight unique ways of engineering this 2-by-2 Prisoner's Dilemma game to transition to a Cooperative (also known as Coordination) game.

**A Snowdrift game with a 2-by-2 payoff matrix**: Next, we investigated another simple prototype game with a 2-by-2 payoff matrix but this time with two cells serving as the NE (**Figure 3a**). This is a typical Snowdrift game where adopting the opposite strategy to the opponent is the NE—specifically, both (C, D) and (D, C) constitute NE outcomes. Assuming that this is an undesired outcome, our objective is to steer away from it and transition to a Cooperative game where (C, C) stands as the NE, aligning with our preferred strategic outcome. For this game, we identified only one unique game engineering strategy that facilitates the transition to a (C, C) NE (**Figure 3b**). This constitutes perturbing two different payoffs in cell (C, C). These include: (i) Increasing the payoff of a cooperating Player 1 when facing a cooperating Player 2 by 2.01. This adjustment makes cooperation a more favorable strategy (compared to defecting) for Player 1 when Player 2 cooperates (payoff of 5.01 for cooperation vs. 5 for defecting). (ii) Increasing the payoff of cooperation for Player 2 when encountering a cooperating Player 1 by 2.01. This incentivizes Player 2 to cooperate (instead of defecting) when Player 1 cooperates (resulting in a payoff of 5.01 for cooperation vs. 5 for defecting). These two interventions both make (C, C) an NE and prevent (C, D) and (D, C) from being NE. Of note, our pipeline did not identify any other alternative strategies to transition to (C, C) since we did not allow any perturbations to cells of the original NE, namely, (C, D) and (D, C), as noted before.

**A two-player game with a 5-by-5 payoff matrix**: Next, we extended our evaluation to a more intricate setting, examining the effectiveness of our pipeline to engineer a two-player game with a 5-by-5 payoff matrix. Here, each of the two players can choose from a set of five different strategies denoted as $S1, S2, …., S5$ as illustrated in **Figure 4a**. In this game, the cell ($S3, S3$) is the NE, which we deem an undesired outcome. We seek to identify targeted changes to the payoff matrix that would enable us to move from this undesired NE to a more favorable equilibrium state, specifically, ($S5, S2$). Utilizing our game engineering pipeline, we identified a total of 38 unique alternative intervention strategies that facilitate transitioning away from ($S3, S3$) and establishing ($S5, S2$) as the NE. **Figures 4b** to **4i** showcase the first eight identified interventions, with a comprehensive list provided in **Supplementary Figure 1**. For example, the game engineering strategy illustrated in **Figure 4b** includes three perturbations: (i) Increasing the payoff of Player 1 when taking strategy $S5$ against Player 2 taking strategy $S2$ by 7.01. This makes $S5$ the most favorable strategy (i.e., the strategy with maximum payoff) for Player 1 when facing Player 2 taking $S2$. (ii) Increasing the payoff of Player 2 taking $S2$ when facing a Player 1 that takes $S5$ by 7.01. This incentivizes Player 2 to take $S2$ when Player 1 takes $S5$, giving Player 2 a maximum payoff of -0.99. Perturbations (i) and (ii) facilitate ($S5, S2$) becoming the new NE. (iii) The last identified perturbation increases the payoff of Player 1 taking $S5$ when facing Player 2 taking $S3$ by 3.01. This motivates Player 1 to take $S5$ when Player 2 chooses $S3$, thus preventing ($S3, S3$) from being an NE. The other engineering strategies shown in **Figures 4c** to **4i** provide alternative ways of reshaping the strategic payoffs to achieve the same goal.

**A four-player game**: Our proposed framework can readily handle games with more than two players. To demonstrate this capability, we analyzed a four-player game where each player has the option to either Cooperate (C) or Defect (D). The strategy space of this game was transformed into a 2-by-2 representation of the payoff matrix as depicted in **Figure 5a** for the sake of illustration. With the specific payoffs shown, all players defecting, i.e., the configuration (D, D, D, D), emerges as the NE—an outcome we aim to transition away from. This can be considered a generalized form of the classical Prisoner's Dilemma game where defection is the dominant strategy for all players. Our objective is to transform this game into a Cooperative (Coordination) game where all players cooperating, i.e., (C, C, C, C), is the NE. Utilizing our game engineering framework, we identified 32 distinct and viable strategies to facilitate this transition (four of which are shown in **Figures 5b** to **5e**), showcasing the framework's versatility in redefining equilibrium states in complex game settings. For example, the game engineering strategy shown in **Figure 5b** involves four interventions: (i) Increasing the payoff of a cooperating Player 1 when facing cooperating Players 2, 3, and 4 by 5.02. This adjustment makes cooperation a more favorable strategy (compared to defecting) for Player 1 when all other players cooperate (payoff of 8.02 for cooperation vs. 8 for defecting). (ii) Increasing the payoff of a cooperating Player 3 when facing cooperating Players 1, 2, and 4 by 1.02. This adjustment makes cooperation a more favorable strategy (compared to defecting) for Player 3 when all other players cooperate (payoff of 3.02 for cooperation vs. 3 for defecting). (iii) Increasing the payoff of a cooperating Player 4 when facing cooperating Players 1, 2, and 3 by

2.02. This incentivizes Player 4 to cooperate when all other players cooperate (payoff of 8.02 for cooperation vs. 8 for defecting). These three interventions make (C, C, C, C) an NE. (iv) Lastly, increasing the payoff of a cooperating Player 4 when Players 1, 2, and 3 defect by 0.02, which makes cooperation a more favorable strategy for Player 4 than defecting (a payoff of 9.02 compared to 9), thereby preventing (D, D, D, D) from being the NE. The details of the 28 remaining identified engineering strategies for this four-player game is given in **Supplementary File 2**.

**Navigating complex multi-strategy games**: In order to evaluate the scalability of our game engineering pipeline for complex and intricate game scenarios, we could explore games with either multiple players or two-player games with multiple strategies. Note that regardless of multi-player or multi-strategy game types, what determines the complexity of the game engineering problem is the number of cells in the payoff matrix. For the ease of presentation and representation, here we chose the latter, i.e., two-player games where each player chooses from an expanding array of different strategies. To this end, we created synthetic two-players games and progressively increased the complexity of the strategy space, where the number of strategies each player can choose from ranged from 20 to 1000—equivalent to 400 to $10^6$ cells in the payoff matrix and 800 to $2 \times 10^6$ payoffs to perturb. For each specified number of strategies available to each player, $n$, we initialized a random payoff matrix of size $n \times n$ in such a way that a randomly chosen cell in the payoff matrix serves as our original undesired NE. Additionally, we designated another random cell in the payoff matrix as the desired NE. We then ran our game engineering pipeline on each game to identify multiple unique alternative strategic interventions enabling the transition from the undesired to the desired NE. Moreover, to make our results more robust, we initialized multiple random games for each game size, $n^2$, and ran our pipelines as described to find multiple alternative interventions for each initialization. This resulted in a range of 17 to 48 interventions for each game size. Recording the CPU times for finding each intervention, we plotted the median runtime over all alternative intervention strategies for each specific game size, $n^2$. As depicted in **Figure 6,** we observe an almost linear trend in the CPU runtime with respect to the number of cells in the payoff matrix, $n^2$ ($R^2 = 0.96$). These observations underscore the capability of our framework to scale for the engineering of fairly complex games.

## Discussion

In this study, we presented an optimization-driven framework for engineering a mathematical game with any number of players and strategies by tweaking the strategic payoffs of the game. This approach ensures convergence to a precisely tailored NE while effectively precluding the emergence of undesired NE states. Our approach revolves around the idea of identifying the most critical players in the game and strategies they choose and applying subtle adjustments to their payoffs to ensure the realization of the preferred equilibrium state and the exclusion of undesired ones. Through an array of rigorous computational experiments, we substantiated the capability of our framework, showcasing its versatility in engineering multi-player and multi-strategy games with a diverse array of desirable properties. Furthermore, our study reveals the scalability of our approach, demonstrating its capability to handle games of increasing complexity with near-linear efficiency. Our proposed framework represents a significant departure from previous methods in that it does not rely on finding analytical solutions such as those employed in variational methods for assessing the controllability of a game [18, 21]. Instead, our study complements these methods by designing precise and targeted strategic interventions using a computational approach (grounded in mathematical optimization), thereby exerting control over the game by shifting its NE from one state to another.

This work opens new horizons for a variety of real-world applications involving strategic interactions, offering a powerful toolset for crafting equilibrium outcomes that meet precise specifications. In the realm of political science, our framework may find applications in guiding the optimization of strategic interactions between international entities, for example, by systematically identifying effective strategies for reshaping negotiations, incentivizing cooperation, and mitigating conflicts. This could lead to the establishment of more stable agreements, the resolution of long-standing disputes, and the fostering of improved international cooperation. In the field of economics, our framework may be employed to meticulously design and fine-tune economic incentives and market strategies that promote fairer market outcomes and mitigate economic volatility. For instance, it could be applied to optimize auction mechanisms for more efficient and fair resource allocation, or to fine-tune market regulations to prevent market abuse while promoting competition and innovation, thereby minimizing market distortions. These applications can help promote a more equitable and robust economic system. In the field of biology, our approach could facilitate the design of strategic interventions, for example, for the preservation of vital ecological systems. Within this context, our game engineering pipeline could serve as a potent tool for crafting precisely calibrated ecological incentives and strategies aimed at preserving biodiversity, e.g., by promoting species coexistence and identifying effective measures to encounter invasive species.

It is important to recognize some limitations of our work. Firstly, in evaluating the scalability of our approach, we assumed there is one undesired NE and one preferred NE, which resulted in an approximately linear runtime relative to the size of the payoff matrix (**Figure 6**). While we anticipate this near-linear relationship to persist in scenarios with a limited number of undesired and desired NE, the linear scaling may not be necessarily sustained in cases with an excessive number of undesired and/or desired NE. Nevertheless, in practical applications, the quantity of undesired and desired NE is typically restricted; therefore, we do not expect this to substantially impact the practical applicability of our framework. Another limitation of our approach is that it focuses solely on pure strategy NE precluding its application to scenarios involving mixed strategy NE, which involve probabilistic decision-making by players. Furthermore, our framework does not encompass the modeling of iterative games, where the strategies employed by players at any given moment are influenced by the strategies deployed by players in previous time points. These constraints confine the utility of our framework to contexts involving strategic interactions characterized by deterministic strategies and static games. The real-world applicability of our results may be influenced by the degree to which these assumptions hold. While these limitations delineate the scope of our approach, they also underscore opportunities for future enhancements and adaptations to accommodate a broader spectrum of strategic interactions.

Overall, our game engineering framework offers a versatile toolkit for precision strategic decision-making with far-reaching implications across diverse domains. As we delve deeper into the synthesis of game theory and practical applications, the horizons for harnessing equilibrium engineering become expansive. Future research endeavors should concentrate on the seamless translation of the equilibrium shift strategies generated by our framework into actionable and context-specific modifications to unleash its full potential in addressing a myriad of real-world challenges.

## Methods

Our game engineering pipeline is based on a mathematical optimization approach. The goal of this optimization problem is to identify minimal perturbations to the payoffs of a game that lead to the change of its pure strategy NE to another state. For the simplicity of our exposition, we present the formulation for a two-player game with an arbitrary number of strategies. Extension to an $n$-player game with any number of strategies is straightforward. Let the payoffs in the $ij$th cell of the payoff matrix of this two-player game be denoted by $(a_{ij}^1, a_{ij}^2), \forall i \in S_1, j \in S_2$, where $S_1$ and $S_2$ are the strategy spaces for players 1 and 2, and $a_{ij}^1$ and $a_{ij}^2$ are the payoffs of players 1 and 2, respectively, when Player 1 chooses strategy $i$ and Player 2 chooses strategy $j$. To model perturbations to the payoff values, we define two non-negative variables $\alpha_{ij}^{k^+}, \alpha_{ij}^{k^-} \geq 0$, $k \in \{\text{Player 1}, \text{Player 2}\}$ and impose the perturbation to the payoffs as follows:

$$b_{ij}^k = a_{ij}^k + \alpha_{ij}^{k^+} - \alpha_{ij}^{k^-}, \qquad \forall\, i \in S_1,\ j \in S_2,\ k \in \{\text{Player 1}, \text{Player 2}\}, \qquad (1)$$

where, $b_{ij}^k$ is the adjusted payoff after perturbation. This definition allows perturbations to payoff values in both directions, i.e., to either increase or decrease.

Let $D$ be the set of all cells of the payoff matrix of the game that serve as the desired NE and $U$ be the set of all cells corresponding to the current undesired NE states. Our goal is to minimally perturb the original payoff matrix to transform it into a new game for which cells in $D$ are the NE while those in $U$ no longer satisfy the conditions of the NE.

**Enforcing the desired NE**: To enforce a cell $rc \in D$ to be an NE, the following constraints are imposed:

$$b_{rc}^1 \geq b_{ic}^1 + \epsilon, \qquad \forall rc \in D, \qquad \forall i \in \{i | i \in S_1\ \&\ i \neq r\}, \qquad (2)$$

$$b_{rc}^2 \geq b_{rj}^2 + \epsilon, \qquad \forall rc \in D, \qquad \forall j \in \{j | j \in S_2\ \&\ j \neq c\}, \qquad (3)$$

where $\epsilon$ is a user-defined small positive scalar (e.g., $10^{-5}$) used to impose strict inequalities. Constraint (2) ensures that the highest payoff across all rows in column $c$ belongs to row $r$. Likewise, constraint (3) ensures that the highest payoff across all columns in row $r$ belongs to column $c$. These two conditions imply that the cell $rc$ is a NE.

**Preventing undesired NE**: To prevent undesired states from being NE in the engineered game, we define the following variables for every undesired cell $rc \in U$: $b_c^{1,max}$ and $b_r^{2,max}$, where $b_c^{1,max}$ represents the maximum payoff of the Player 1 in the perturbed payoff matrix when Player 2 takes the strategy corresponding to $c$ ($b_c^{1,max} = \max_{i \in S_1} b_{ic}^1$), and $b_r^{2,max}$ is the maximum payoff of Player 2 in the perturbed payoff matrix when Player 1 takes the strategy corresponding to $r$ ($b_r^{2,max} = \max_{j \in S_2} b_{rj}^2$). In order to prevent the cell $rc$ from being an NE, it is sufficient to have either $b_{rc}^1 < b_c^{1,max}$ or $b_{rc}^2 < b_r^{2,max}$. We can impose these conditions using the following constraints:

$$b_c^{1,max} \geq b_{rc}^1 + \epsilon, \qquad \forall rc \in U, \tag{4}$$

$$b_c^{1,max} \geq b_{ic}^1, \qquad \forall c \in \{c|rc \in U\}, \quad i \in S_1 \tag{5}$$

$$b_c^{1,max} \leq b_{ic}^1 + (1-w_{ic}^1)M, \qquad \forall c \in \{c|rc \in U\}, \quad \forall i \in \{i|i \in S_1 \ \& \ i \neq r\}, \tag{6}$$

$$\sum_{i \in S_1} w_{ic}^1 \leq 1 \qquad \forall c \in \{c|rc \in U\}, \tag{7}$$

$$b_r^{2,max} \geq b_{rc}^2 + \epsilon, \qquad \forall rc \in U \tag{8}$$

$$b_r^{2,max} \geq b_{rj}^2, \qquad \forall r \in \{r|rc \in U\}, \quad j \in S_2 \tag{9}$$

$$b_r^{2,max} \leq b_{rj}^2 + (1-w_{rj}^2)M, \qquad \forall r \in \{r|rc \in U\}, \quad \forall j \in \{j|j \in S_2 \ \& \ j \neq c\}, \tag{10}$$

$$\sum_{j \in S_2} w_{rj}^2 \leq 1 \qquad \forall r \in \{r|rc \in U\}, \tag{11}$$

$$\sum_{i \in S_1} w_{ic}^1 + \sum_{j \in S_2} w_{rj}^2 \geq 1, \qquad \forall rc \in U, \tag{12}$$

$$w_{ic}^1 \in \{0,1\}, \qquad \forall c \in \{c|rc \in U\}, \quad i \in \{i|i \in S_1 \ \& \ i \neq r\},$$

$$w_{rj}^2 \in \{0,1\}, \qquad \forall r \in \{r|rc \in U\}, \quad j \in \{j|j \in S_2 \ \& \ j \neq c\}.$$

Here, $\epsilon$ is a small positive scalar and $M$ is a large positive scalar ("big M"). Constraint (4) imposes $b_{rc}^1 < b_c^{1,max}$. A binary variable $w_{ic}^1$ is defined for each row $i$ ($i \neq r$) in column $c$ indicating whether the $i$th entry in this column corresponds to the maximum payoff of Player 1.

Constraint (5) imposes $b_c^{1,max} = \max\limits_{i \in S_1} b_{ic}^1$. Constraint (6) sets $b_c^{1,max}$ to $b_{ic}^1$, if $w_{ic}^1 = 1$. Constraint (7) requires at most one row in columns $c$ (other than row $r$) to have the maximum payoff (i.e., no ties). Notice that if all $w_{ic}^1$'s are zero, $b_c^{1,max}$ can take any value less than $M$, and Constraint (4) is no longer binding.

Constraints (8) to (11) are similar to Constraints (4) to (7). More specifically, Constraint (8) imposes $b_{rc}^2 < b_r^{2,max}$. The binary variable $w_{rj}^2$ is defined for each column $j$ ($j \neq c$) in row $r$ indicating whether the $j$th entry in this row corresponds to the maximum payoff of Player 2, Constraint (9) imposes $b_r^{2,max} = \max\limits_{j \in S_2} b_{rj}^2$ and Constraint (10) sets $b_r^{2,max}$ to $b_{rj}^2$, if $w_{rj}^2 = 1$. Constraint (11) requires at most one column in row $r$ to have the maximum payoff. Note that the requirement for having no ties in the maximum payoffs imposed in Constraints (7) and (11) does not compromise the generality of our framework as it can be easily relaxed. Constraint (12) ensures that at least one of the two conditions $b_{rc}^1 < b_c^{1,max}$ or $b_{rc}^2 < b_r^{2,max}$ is imposed.

**Game engineering optimization formulation**: The optimization problem to pinpoint strategic interventions within the game that facilitate the shift from the current NE to a more preferred NE is as follows:

$$\text{Minimize} \sum_{i \in S_1} \sum_{j \in S_2} \sum_{k \in \{\text{Player 1}, \text{Player 2}\}} \left( \alpha_{ij}^{k^+} + \alpha_{ij}^{k^-} \right)$$

subject to

$$b_{ij}^k = a_{ij}^k + \alpha_{ij}^{k^+} - \alpha_{ij}^{k^-}, \quad \forall\, i \in S_1,\ j \in S_2,\ k \in \{\text{Player 1}, \text{Player 2}\}, \tag{1}$$

$$b_{rc}^1 \geq b_{ic}^1 + \epsilon, \quad \forall (r,c) \in D, \quad \forall i \in \{i | i \in S_1 \ \&\ i \neq r\}, \tag{2}$$

$$b_{rc}^2 \geq b_{rj}^2 + \epsilon, \quad \forall (r,c) \in D, \quad \forall j \in \{j | j \in S_2 \ \&\ j \neq c\}, \tag{3}$$

$$b_c^{1,max} \geq b_{rc}^1 + \epsilon, \quad \forall rc \in U \tag{4}$$

$$b_c^{1,max} \geq b_{ic}^1, \quad \forall c \in \{c | rc \in U\}, \quad i \in S_1 \tag{5}$$

$$b_c^{1,max} \leq b_{ic}^1 + (1 - w_{ic}^1)M, \quad \forall c \in \{c | rc \in U\}, \quad \forall i \in \{i | i \in S_1 \ \&\ i \neq r\}, \tag{6}$$

$$\sum_{i \in S_1} w_{ic}^1 \leq 1 \quad \forall c \in \{c | rc \in U\}, \tag{7}$$

$$b_r^{2,max} \geq b_{rc}^2 + \epsilon, \quad \forall rc \in U \tag{8}$$

$$b_r^{2,max} \geq b_{rj}^2, \qquad \forall r \in \{r|rc \in U\}, \quad j \in S_2 \qquad (9)$$

$$b_r^{2,max} \leq b_{rj}^2 + (1-w_{rj}^2)M, \qquad \forall r \in \{r|rc \in U\}, \quad \forall j \in \{j|j \in S_2 \,\&\, j \neq c\}, \qquad (10)$$

$$\sum_{j \in S_2} w_{rj}^2 \leq 1 \qquad \forall r \in \{r|rc \in U\}, \qquad (11)$$

$$\sum_{i \in S_1} w_{ic}^1 + \sum_{j \in S_2} w_{rj}^2 \geq 1, \qquad \forall rc \in U, \qquad (12)$$

$$\alpha_{ij}^{k\,+}, \alpha_{ij}^{k\,-} \geq 0, \qquad \forall i \in S_1, \; j \in S_2, \; k \in \{\text{Player 1, Player 2}\},$$

$$w_{ic}^1 \in \{0,1\}, \qquad \forall c \in \{c|rc \in U\}, \quad i \in \{i\,|\,i \in S_1 \,\&\, i \neq r\},$$

$$w_{rj}^2 \in \{0,1\}, \qquad \forall r \in \{r|rc \in U\}, \quad j \in \{j\,|\,j \in S_2 \,\&\, j \neq c\}.$$

The objective function of this optimization problem minimizes total perturbations to the payoff matrix of the game. The constraints are as described before and collectively enforce the desired NE while preventing the undesired ones. Note that since we prefer not to perturb the payoffs in the cells corresponding to current undesired NE, we can set $\alpha_{rc}^+ = 0$ and $\alpha_{rc}^- = 0$ for all $rc \in U$.

**Finding alternative solutions**: Here, we describe how one can find alternative solutions to the optimization problem above, i.e., alternative ways of perturbing the payoff matrix to achieve the same desired NE while precluding any undesired NE. Our goal is finding unique alternative interventions that do not involve perturbing the exact same combination of the payoffs identified in any previous solution. To this end, we define a binary variable $y_\alpha$ for each perturbation value, $\alpha$ (where $\alpha$ can be any $\alpha_{ij}^{k\,+}$ or $\alpha_{ij}^{k\,-}$), that assumed a non-zero optimal value in any of the previous solutions as follows:

$$y_\alpha = \begin{cases} 1 & \text{if } \alpha \neq 0 \\ 0 & \text{otherwise,} \end{cases} \quad \forall \alpha \in \{\alpha | \alpha \neq 0 \text{ in any previous solution}\}.$$

This definition can be imposed using the following constraint:

$$\epsilon y_\alpha + (-M)(1 - y_\alpha) \leq \alpha \leq (+M) y_\alpha \qquad \forall \alpha \in \{\alpha | \alpha \neq 0 \text{ in any previous solution}\}, \qquad (13)$$

where $\epsilon$ is a small positive scalar. Note that this constraint together with the non-negativity constraints on $\alpha$ (i.e., $\alpha \geq 0$) ensure that $y_\alpha = 1 \leftrightarrow \alpha \neq 0$ and $y_\alpha = 0 \leftrightarrow \alpha = 0$.

When finding the next alternative solution, we need at least one perturbed payoff (i.e., at least one entry in the payoff matrix) to be different from the those we had found the previous solutions. We can impose this by using integer cuts described with the following constraint:

$$\sum_{\alpha \in NZ} y_\alpha \leq (card(NZ_\alpha) - 1) \tag{14}$$

where $NZ_\alpha$ is the set of all $\alpha$'s that were non-zero in the previous solution and $card(NZ_\alpha)$ is cardinality of this set. This constraint should be added for every solution found in previous iterations.

**Computational implementation**: All computational simulations were conducted in Python 3. Optimization problems were implemented using the Optlang Python package (https://optlang.readthedocs.io/en/latest/) and were solved using the Gurobi solver (Gurobi Optimization, LLC).

## Supplemental information
**Supplementary Figure 1**: The comprehensive list of solutions for the two-player game with 5-by-5 payoff matrix.
**Supplementary File 1**. A sample implementation of prototype games presented in this article as Google Collaboratory notebooks.
**Supplementary File 2**: The comprehensive list of the identified engineering strategies for the four-player game.

## Code availability
All the source code is available on GitHub (https://github.com/zomorrodilab/game_engineering) and additionally as part of the Google Collaboratory notebooks in **Supplementary File 1**.

## Author contributions
ARZ conceived the study. ES performed all implementations and computational analyses. Both authors contributed to writing and editing the manuscript.

Figures

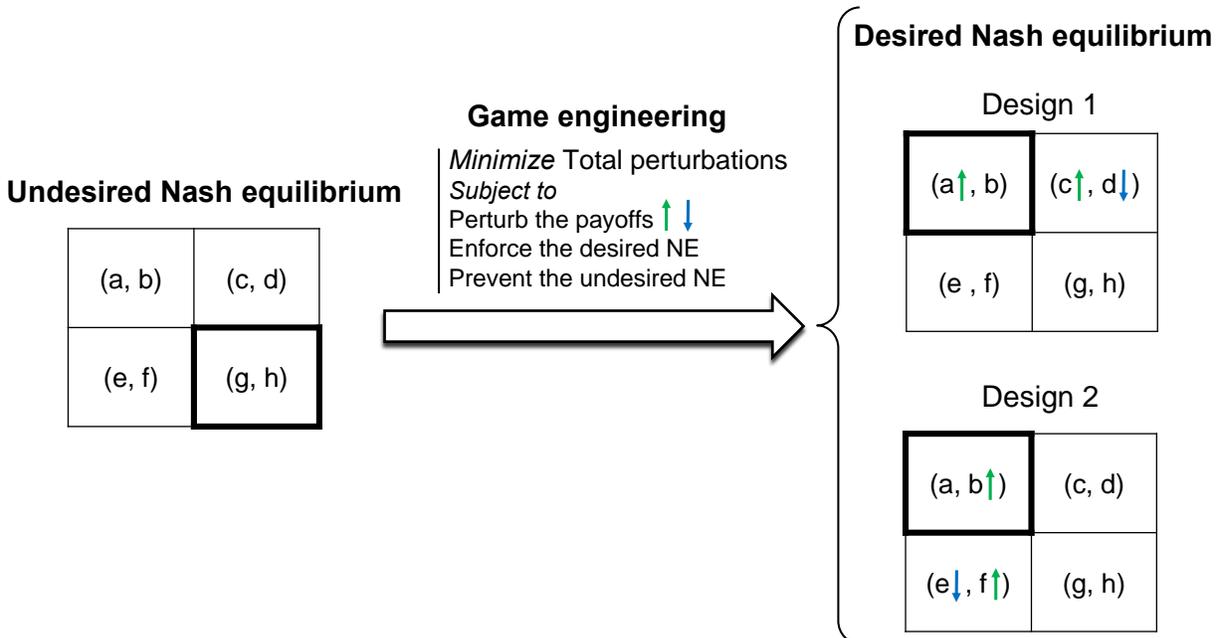

**Figure 1**. A schematic overview of the proposed game engineering framework. This pipeline pinpoints the intricate combination of specific players and strategic payoffs and the optimal magnitude of perturbations to these payoffs. Crucially, the framework achieves this while minimizing the overall perturbation to the system to enable the transition from an undesired NE to a desired one.

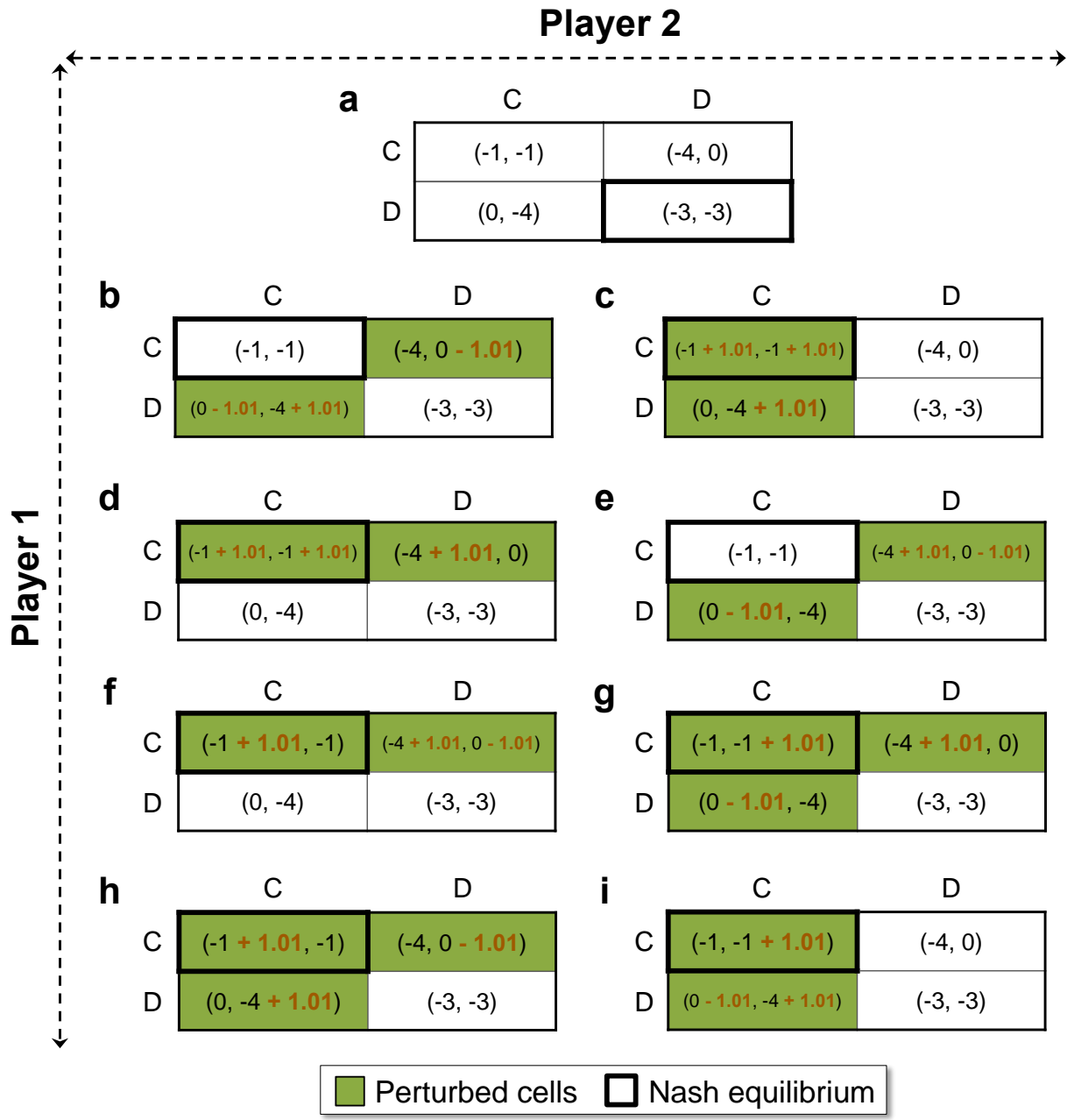

Figure 2. All alternative ways of perturbing a typical Prisoner's Dilemma game with a 2-by-2 payoff matrix (panel **a**) to transform it into a Cooperative (Coordination) game (panels **b** to **i**).

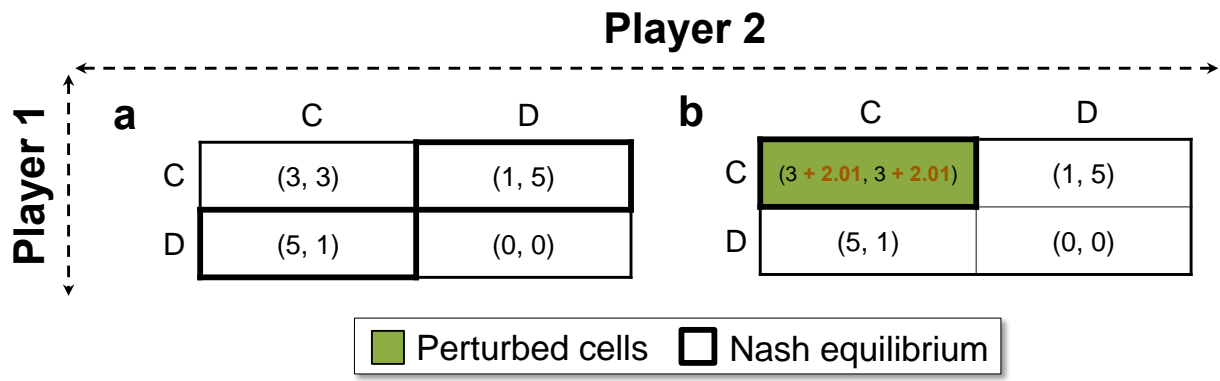

**Figure 3**. Perturbing a standard Snowdrift game with a 2-by-2 payoff matrix (panel **a**) to transform it into a Cooperative (Coordination) game (panel **b**).

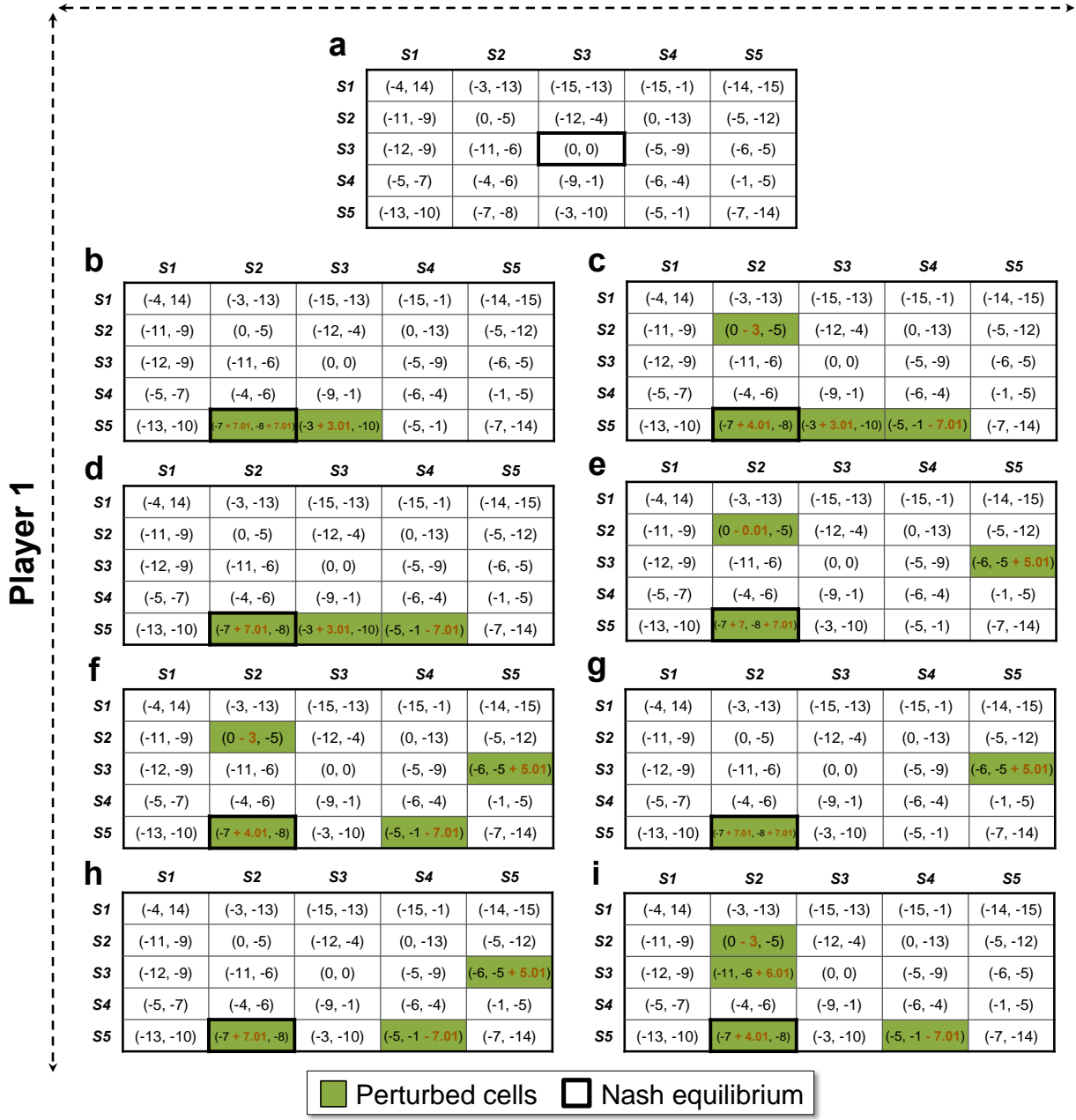

Figure 4. Eight (out of the 38) unique ways of engineering a game with a 5-by-5 payoff matrix (panel **a**) in order to change the NE from ($S3, S3$) to ($S5, S2$) (panels **b** to **i**).

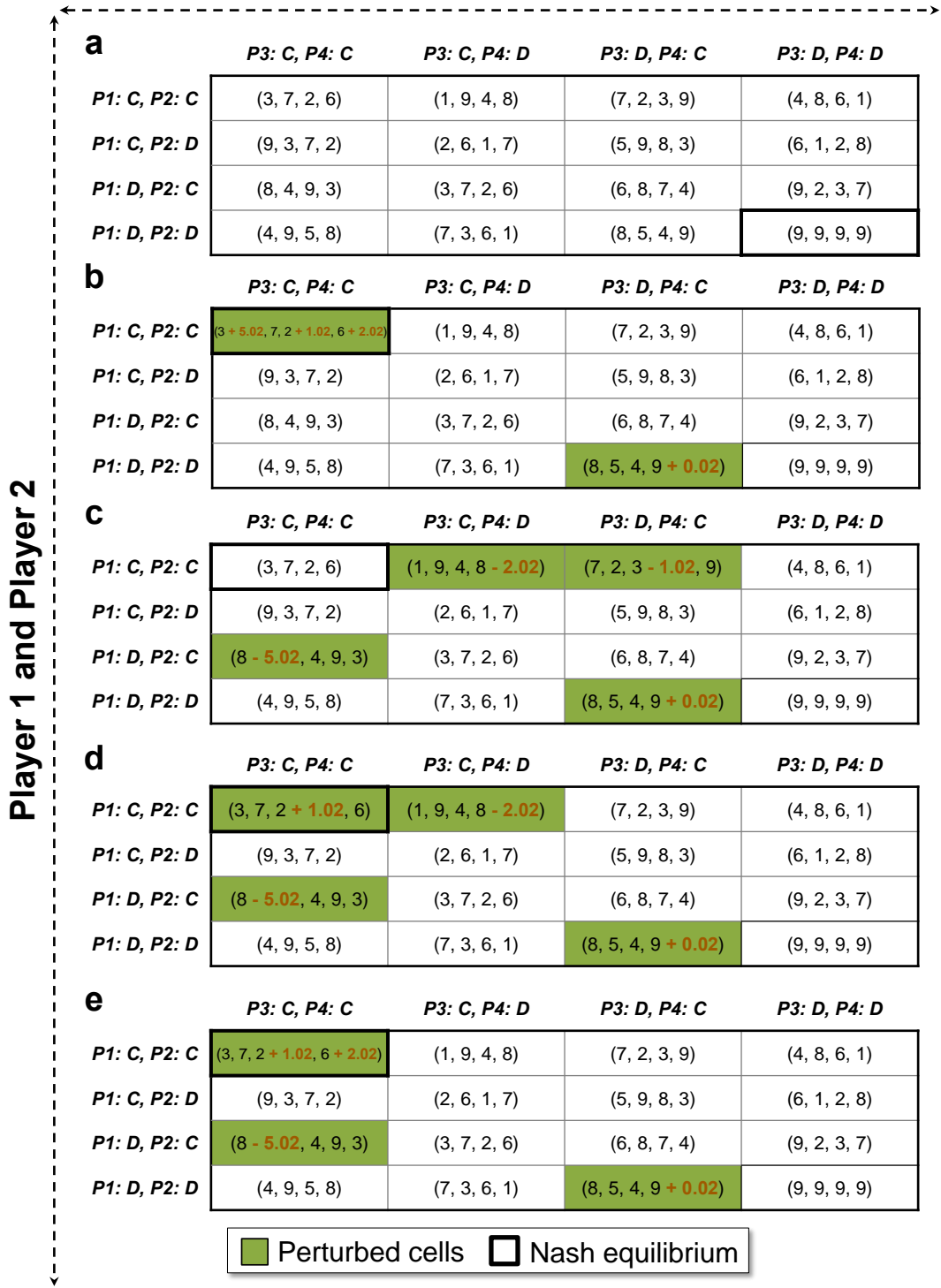

**Figure 5.** Four (out of the 32) unique ways of engineering a four-player game. The objective here is to change the NE from (D, D, D, D) (panel **a**) to (C, C, C, C) (panels **b** to **e**).

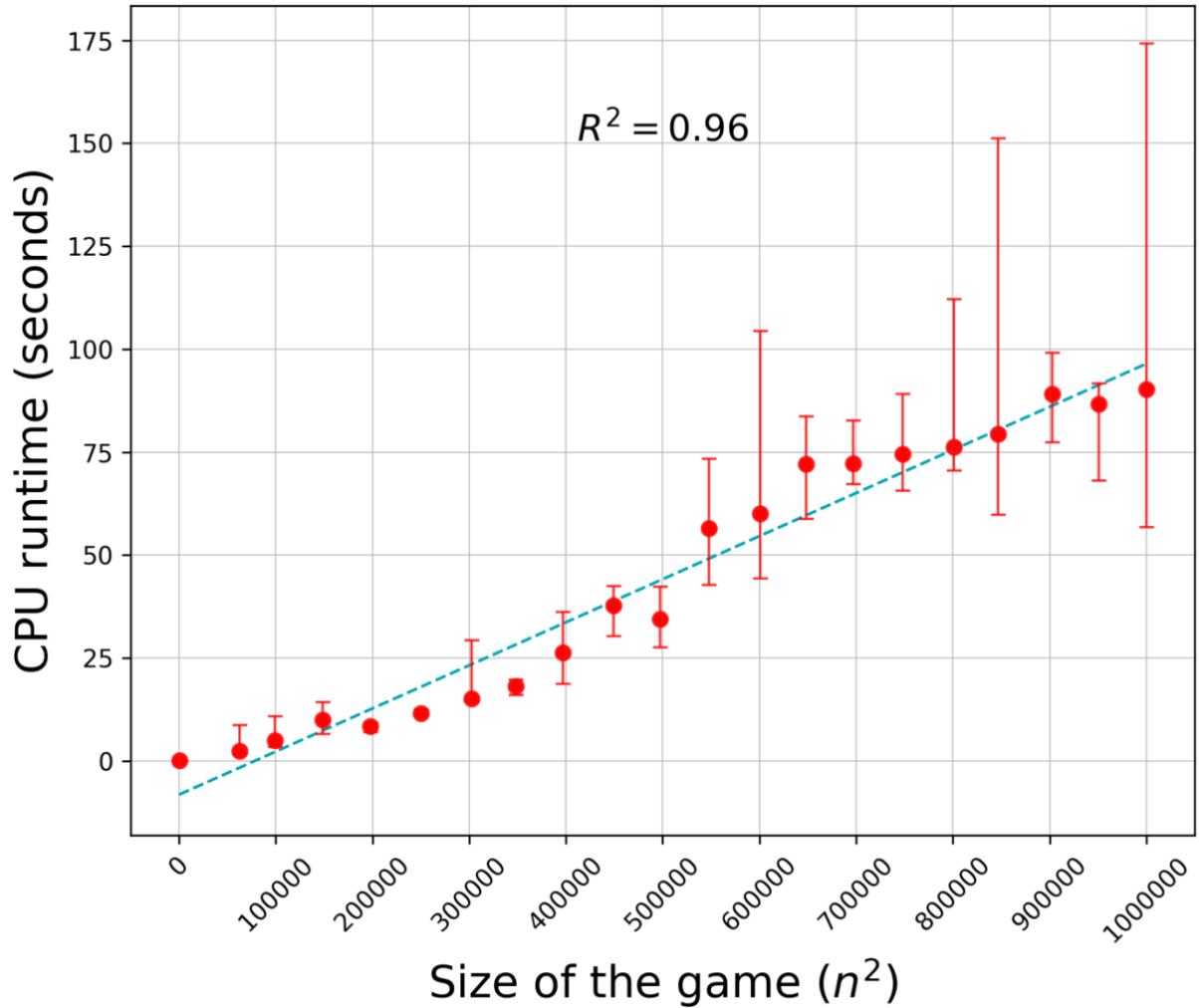

**Figure 6**. Median CPU time for identifying game engineering interventions for games of increasing complexity. The median CPU time is reported for identifying perturbations to two-player games relative to the size of the strategy space, represented by the number of entries in the payoff matrix, $n \times n$. The median was derived across multiple runs, ranging from 17 to 48 runs, for finding alternative solutions for a given game. The runtimes are reported for analysis conducted on a compute cluster node with Intel(R) Xeon(R) Gold 6140 CPUs at 2.30GHz.